\documentclass[11pt]{article}

\usepackage{authblk}
\usepackage[colorlinks,linkcolor=blue,anchorcolor=red,citecolor=red,pdfstartview=FitH]{hyperref}
\usepackage{graphics,graphicx}
\usepackage{mathtools}
\usepackage{amsmath,amssymb,amsthm,accents,amsfonts}
\usepackage{color}
\usepackage{cite}
\usepackage{fullpage}

\DeclareMathAlphabet{\mathpzc}{OT1}{pzc}{m}{it}
\DeclareMathOperator{\antidiag}{\mathrm{antidiag}}
\newtheorem{theorem}{Theorem}[section]
\numberwithin{equation}{section}
\allowdisplaybreaks

\begin{document}

\title{\bf {Exact solutions of the Gerdjikov-Ivanov equation using Darboux transformations}}
\author[1,2]{\bf Halis Yilmaz\footnote{E-mail: Halis.Yilmaz@glasgow.ac.uk, halisyilmaz@dicle.edu.tr}}

\affil[1]{School of Mathematics and Statistics, University of Glasgow, Glasgow G12 8QW, UK}
\affil[2]{Department of Mathematics, University of Dicle, 21280 Diyarbakir, Turkey}
\date{\today}
\maketitle

\begin{abstract}
 We study the Gerdjikov-Ivanov (GI) equation and present a standard Darboux transformation for it. The solution is given in terms of quasideterminants. Further, the parabolic, soliton and breather solutions of the GI equation are given as explicit examples.
\end{abstract}

\quad{\text{\it{Keywords:}} Gerdjikov-Ivanov equation; Derivative nonlinear Schr\"{o}dinger equation; Darboux transformation;  Quasideterminants.}

\quad{\text{\it{2010 Mathematics Subject Classification:}} 35C08, 35Q55, 37K10, 37K35}

\section{Introduction}
The well known nonlinear Schr\"{o}dinger (NLS) equation is one of the most important soliton equations.
Extended versions of this equation with higher order nonlinearity have been proposed and studied by various authors.
Among them, there are three celebrated equations with derivative--type nonlinearities, which are called the derivative NLS equations (DNLS).
One is the Kaup-Newell equation (DNLSI) \cite{KN}
\begin{eqnarray*}
 iq_t+q_{xx}=i(|q|^2q)_x,
\end{eqnarray*}
the second is the Chen--Lee--Liu equation (DNLSII) \cite{Chen}
\begin{eqnarray*}
 iq_t+q_{xx}+i|q|^2q_x=0,
\end{eqnarray*}
while the third is the Gerdjikov--Ivanov equation (DNLSIII) \cite{GI}
\begin{eqnarray}\label{GIeqn}
 iq_t+q_{xx}+ iq^2q^*_x+\frac{1}{2}q^3q^{*^2}=0,
\end{eqnarray}
where $q^*$ denotes the complex conjugate of $q$.
The NLS equation  with its cousin the DNLS equations are completely integrable and play an important role in mathematical physics \cite{Agr,Clarkson,Ich,Johnson,Kodama,Ohta}. 

It is known that these three equations may be transformed into each other by a chain of gauge transformations and the method of gauge transformation can also be applied to some generalised cases \cite{Clarkson87,Kakei,Kundu84,Kundu,Wadati}. Therefore, in principle, the corresponding results for Chen-Lee--Lue and the GI equations may be obtained from the Kaup-Newell equation. However, these transformations involve very complicated integrals and it is not easy to obtain their explicit  forms. So, even though the three systems are related by gauge transformations it is more convenient to treat them each separately.

In \cite{NH}, the explicit quasideterminant solutions of the Kaup Newell equation (DNLSI) are presented via a standard Darboux transformation. In this paper, we study the Gerdjokov-Ivanov equation (DNLSIII) to obtain explicit solutions by using a standard Darboux transformation. Darboux  transformations are an important tool for studying the solutions of integrable systems. They provide a universal algorithmic procedure to derive explicit exact solutions of integrable systems. In recent years, there has been some interest in solutions of  the Gerdjikov-Ivanov equation obtained by means of \emph{Darboux-like} transformations \cite{Fan,Guo,Xu}. These solutions are often written in terms of determinants with a complicated structure, where the determinant representations of $n$-fold Darboux transformations are obtained by stating and proving a sequence of theorems.

On the other hand, in the present paper, we present a systematic approach to the construction of solutions of \eqref{GIeqn} by means of a standard Darboux transformation and written in terms of quasideterminants \cite{Gelfand91,Gelfand05}. Quasideterminants have various nice properties which play important roles in constructing exact solutions of integrable systems \cite{CJ,GNH,Hassan,Nimmo06,NH,NGO}.

This paper is organized as follows. In Section \ref{Qdet} below, we give a brief review on quasideterminants.
In Section \ref{DTDR}, we state a standard Darboux theorem for the Gerdjikov-Ivanov system. In Sections \ref{DT1} and \ref{DT2}, we present the quasideterminant solutions of the Gerdjikov-Ivanov equation by using the Darboux transformation. Here, the quasideterminants are written in terms of solutions of linear eigenvalue problems. In Section \ref{ParS}, particular solutions of the Gerdjikov-Ivanov equation are given for both zero and non-zero seed solutions. The conclusion is given in the final Section \ref{Con}.

\subsection{Quasideterminants}\label{Qdet}
In this short section we recall some of the key elementary properties of quasideterminants. The reader is referred to the original papers \cite{Gelfand91,Gelfand05} for a more detailed and general treatment.

The notion of a quasideterminant was first introduced by Gelfand  and Retakh in \cite{Gelfand91} as a straightforward way to define the determinant of a matrix with noncommutative entries. Many equivalent definitions of quasideterminants exist, one such being a recursive definition involving inverse minors. Let $A=(a_{ij})$ be an $n\times n$ matrix  with entries over a usually non commutative ring 
\begin{equation}\label{expan}
 |A|_{ij}=a_{ij}-r_i^j\left(A^{ij}\right)^{-1}c_j^i ,
\end{equation}
where $r_i^j$ represents the row vector obtained from $i^{th}$ row of $A$ with the $j^{th}$ element removed, $c_j^i$ represents the column vector obtained from $j^{th}$ column of $A$ with the $i^{th}$ element removed and $A^{ij}$ is the $(n-1)\times (n-1)$ submatrix obtained by deleting the $i^{th}$ row and the $j^{th}$ column from $A$.
Quasideterminants can also be denoted by boxing the entry about which the expansion is made
\begin{align}
 |A|_{ij}=
         {\left|\begin{array}{cc}
          A^{ij}& c_j^i\\ r_i^j & \boxed{a_{ij}}
         \end{array}\right|}.
\end{align}
If $A$ is an $n \times n$ matrix over a commutative ring, then the quasideterminant $|A|_{ij}$ reduces to a ratio of determinants
\begin{equation}\label{comdet}
 |A|_{ij}=(-1)^{i+j}\frac{\text{det} A}{\text{det} A^{ij}}.
\end{equation}
It should be noted that the expansion formula \eqref{expan} is also valid in the case of block matrices provided the matrix to be inverted is square. 

In this paper, we will consider only quasideterminants that are expanded about a term in the last column, most usually the last entry. For example considering a block matrix $M=\left(\begin{array}{cc} A& B\\ C & d \end{array}\right)$, where $A$ is an invertible square matrix over $\mathpzc{R}$ of arbitrary size and $B$, $C$ are column and row vectors 
over $\mathpzc{R}$ of compatible lengths, respectively, and $d\in\mathpzc{R}$, the quasideterminant of $M$ is expanded about $d$  is defined by
\begin{align}
         \left|\begin{array}{cc}
          A& B\\ C & \boxed{d}
         \end{array}\right|
    =d-CA^{-1}B.
\end{align}

\section{Gerdjikov-Ivanov equations}\label{GI}
Let us consider the pair of Gerdjikov-Ivanov equations
\begin{eqnarray}
 iq_t+q_{xx}+iq^2r_x+\frac{1}{2}q^3r^2&=&0, \label{GI1}\\
 ir_t-r_{xx}+ir^2q_x-\frac{1}{2}q^2r^3&=&0, \label{GI2}
\end{eqnarray}
where $q=q(x,t)$ and $r=r(x,t)$ are complex valued functions.
Equations (\ref{GI1}) and (\ref{GI2}) reduce to the Gerdjikov-Ivanov equation \eqref{GIeqn} for $r=q^*$ while the choice of $r=-q^*$
would lead to \eqref{GIeqn} with the sign of the nonlinear term reversed.

The Lax pair for the Gerjiov-Ivanov system (\ref{GI1})--(\ref{GI2}) is given by
\begin{eqnarray}
 L&=&\partial_x+J\lambda^2-R\lambda+\frac{1}{2}qrJ\label{LaxL}\\
 M&=&\partial_t+2J\lambda^4-2R\lambda^3+qrJ\lambda^2+U\lambda+W,\label{LaxM}
\end{eqnarray}
where $J$, $R$ and $U$ are $2\times 2$ matrices such that
\begin{eqnarray}
  J={\left(\begin{array}{cc}
          i & 0\\ 0 & -i
  \end{array}\right)},\hspace{0.3cm}
   R={\left(\begin{array}{cc}
          0& q\\ r & 0
  \end{array}\right)},\hspace{0.3cm}
   U={\left(\begin{array}{cc}
          0& -iq_x\\ ir_x & 0
  \end{array}\right)}
\end{eqnarray}
and
\begin{eqnarray}
 W=\left(\begin{array}{cc}
          -\frac{1}{2}\left(rq_x-qr_x\right)-\frac{1}{4}iq^2r^2 & 0\\
             0 & \frac{1}{2}\left(rq_x-qr_x\right)+\frac{1}{4}iq^2r^2
  \end{array}\right).
\end{eqnarray}
Here $\lambda$ is an arbitrary complex number called the eigenvalue (or spectral parameter).

\section{Darboux Theorem and Dimensional Reductions}\label{DTDR}
\begin{theorem}[\cite{Darboux,Matveev,MS}] Consider the linear operator
\begin{equation}\label{L}
 L=\partial_x+\sum_{i=0}^nu_i\partial_y^i
\end{equation}
where $u_i\in R$, where $R$ is a ring, in general non-commutative. Let $G=\theta\partial_y\theta^{-1}$, where $\theta=\theta(x,y)$ is an invertible eigenfunction of $L$, so that $L(\theta)=0$.
Then
\begin{equation}
 \tilde{L}=GLG^{-1}
\end{equation}
has the same form as $L$:
\begin{equation}
 \tilde{L}=\partial_x+\sum_{i=0}^n\tilde{u}_i\partial_y^i
\end{equation}
If $\phi$ is any eigenfunction of $L$ then
\begin{equation}
 \tilde{\phi}=\phi_x-\theta_y\theta^{-1}\phi
\end{equation}
is an eigenfunction of $\tilde{L}$. In other words, if $L(\phi)=0$ then $\tilde{L}(\tilde{\phi})=0$ where $\tilde{\phi}=G(\phi)$.
\end{theorem}

\subsection{Dimensional reduction of Darboux transformation}\label{DRDT}
Here, we describe a reduction of the Darboux transformation from $(2+1)$ to $(1+1)$ dimensions. We choose to  eliminate the $y$-dependence by employing a `separation of variables' technique. The reader is referred to the paper \cite{NGO} for a more detailed treatment. We make the ansatz
\begin{eqnarray}
 \phi &=&\phi^r(x,t)e^{\lambda y},\\
 \theta &=&\theta^r(x,t)e^{\Lambda y},
\end{eqnarray}
where $\lambda$ is a constant scalar and $\Lambda$ an $N \times N$ constant matrix and
the superscript $r$ denotes reduced functions, independent of $y$.
Hence in the dimensional reduction we obtain $\partial_y^{i}\left(\phi\right)=\lambda^i\phi$ and
$\partial_y^{i}\left(\theta \right)=\theta \Lambda^i$ and so the operator $L$ and Darboux transformation $G$ become
 \begin{eqnarray}\label{redL}
  L^r&=&\partial_x+\sum_{i=0}^n u_i\lambda^i,\\
  G^r&=&\lambda-\theta^r\Lambda(\theta^{r})^{-1},
 \end{eqnarray}
where $\theta^r$ is a matrix eigenfunction of $L^r$ such that $L^r\left(\theta^r\right)=0$, with $\lambda$ replaced by the matrix $\Lambda$, that is,
\begin{equation}
	\theta^r_x+\sum_{i=0}^n u_i\theta^r\Lambda^i=0.
\end{equation}
Below we omit the superscript $r$ for ease of notation.

\subsection{Iteration of reduced Darboux Transformations}\label{DT1}

In this section we shall consider iteration of the Darboux transformation and find closed form expressions for these in terms of quasideterminants. 

Let $L$ be an operator, form invariant under the reduced Darboux transformation $G=\lambda-\theta \Lambda \theta^{-1}$ discussed above.

Let $\phi=\phi(x,t)$ be a general eigenfunction of $L$ such that $L(\phi)=0$. Then
\begin{eqnarray*}
\tilde{\phi}&=&G_\theta\left(\phi\right)\\
             &=&\lambda\phi-\theta\Lambda\theta^{-1}\phi\\
             &=&\left|\begin{array}{cc} \theta & \phi\\ \theta\Lambda & \boxed{\lambda\phi}\end{array}\right|
\end{eqnarray*}
is an eigenfunction of $\tilde{L}=G_\theta L G_\theta^{-1}$ so that
$\tilde{L}(\tilde{\phi})=\lambda\tilde{\phi}$. Let $\theta_i$ for $i=1,\ldots,n,$ be a particular set
of invertible eigenfunctions of $L$ so that $L(\theta_i)=0$ for $\lambda=\Lambda_i$, and introduce the notation $\Theta=(\theta_1,\ldots,\theta_n)$. To apply the Darboux transformation a second time, let $\theta_{[1]}=\theta_1$ and $\phi_{[1]}=\phi$ be a general eigenfunction of $L_{[1]}=L$.
Then
$\phi_{[2]}=G_{\theta_{[1]}}\left(\phi_{[1]}\right)$ and $\theta_{[2]}=\phi_{[2]}|_{\phi\rightarrow \theta_2}$ are eigenfunctions for $L_{[2]}=G_{\theta_{[1]}}L_{[1]}G_{\theta_{[1]}}^{-1}$.

In general, for $n\geq 1$, we define the $n$th Darboux transform of $\phi$ by
\begin{equation}
 \phi_{[n+1]}=\lambda\phi_{[n]}-\theta_{[n]}\Lambda_n\theta_{[n]}^{-1}\phi_{[n]},
\end{equation}
in which
\begin{equation*}
 \theta_{[k]}=\phi_{[k]}|_{\phi\rightarrow\theta_k}~.
\end{equation*}
For example,
\begin{eqnarray*}
\phi_{[2]}&=&\lambda\phi-\theta_1\Lambda_1\theta_1^{-1}\phi
        =\left|\begin{array}{cc} \theta_1 & \phi\\ \theta_1\Lambda_1& \boxed{\lambda\phi}\end{array}\right|,\\
\phi_{[3]}&=&\lambda\phi_{[2]}-\theta_{[2]}\Lambda_2\theta_{[2]}^{-1}\phi_{[2]}\\
          &=&\left|\begin{array}{ccc} \theta_1 & \theta_2 & \phi\\
            \theta_1\Lambda_1 & \theta_2 \Lambda_2 & \lambda \phi\\
            \theta_1\Lambda_1^2 &\theta_2\Lambda_2^2 & \boxed{\lambda^2\phi} \end{array}\right|.
\end{eqnarray*}
After $n$ iterations, we get
\begin{eqnarray}
\phi_{[n+1]}=\left|\begin{array}{ccccc}
      \theta_1 & \theta_2 \hspace{0.2cm} \ldots \hspace{0.2cm} \theta_n & \phi\\
      \theta_1\Lambda_1 & \theta_2\Lambda_2 \ldots \theta_n\Lambda_n & \lambda \phi\\
      \theta_1\Lambda_1^2 & \theta_2\Lambda_2^2 \ldots \theta_n\Lambda_n^2 & \lambda^2 \phi\\
      \vdots & \vdots \hspace{0.3cm} \ldots \hspace{0.3cm} \vdots & \vdots \\
      \theta_1\Lambda_1^n & \theta_2\Lambda_2^n \ldots \theta_n\Lambda_n^n & \boxed{\lambda^n\phi}\\
     \end{array}
 \right|.
\end{eqnarray}

\section{Constructing Solutions for Gerdjikov-Ivanov Equation}\label{DT2}
In this section we determine the specific effect of the Darboux transformation $G=\lambda-\theta \Lambda \theta^{-1}$
on the $2\times2$ Lax operators $L,M$ given by \eqref{LaxL},\eqref{LaxM}. Here $\theta$ is a eigenfunction satisfying $L(\theta)=M(\theta)=0$ with $2\times2$ matrix eigenvalue $\Lambda$. By supposing that $L$ is transformed to a new operator $\tilde{L}$, say, we calculate that the effect of the Darboux transformation $\tilde{L}=GLG^{-1}$ is such that
\begin{eqnarray}
 \tilde{R}=R-\left[J,\theta \Lambda \theta^{-1}\right]\label{L1}
\end{eqnarray}
and
\begin{eqnarray}
  \tilde{R}\theta \Lambda \theta^{-1}-\theta \Lambda \theta^{-1} R+\frac{1}{2}J\left(\tilde{q}\tilde{r}-qr\right)&=&0,\label{L2}\\
  \left(\theta \Lambda \theta^{-1}\right)_x+\frac{1}{2}\left[J\left(\theta \Lambda \theta^{-1}\right)\tilde{q}\tilde{r}-\theta \Lambda \theta^{-1}Jqr\right]&=&0.\label{L3}
\end{eqnarray}
From \eqref{L2}, we see that $\theta \Lambda \theta^{-1}$ must be an anti-diagonal matrix, $\antidiag(a, b)$, say, and then from \eqref{L3} the multiplication of the anti-diagonal terms must be constant $(ab=constant)$. Guided by this, we choose
\begin{eqnarray}
 \Lambda=\left(\begin{array}{cc}
        1 & 0\\
             0 & -1
  \end{array}\right)\lambda.
\end{eqnarray}

Finally, the condition $\theta \Lambda \theta^{-1}=\antidiag(a, b)$ leads to the requirement that the matrix $\theta$ has the structure
\begin{eqnarray}
 \theta=\left(\begin{array}{cc}
        \theta_{11} & \theta_{12}\\
             \theta_{21} & \theta_{22}
  \end{array}\right),
\end{eqnarray}
where $\theta_{11}\theta_{22}+\theta_{12}\theta_{21}=0$.

For notational convenience, we introduce a $2\times 2$ matrix $P=(p_{ij})~(i, j=1, 2)$ such that $R=[J,P]$, and hence
\begin{eqnarray}\label{qr}
 P=\frac{1}{2i}\left(\begin{array}{cc}
        p_{11} & q\\
             -r & p_{22}
  \end{array}\right).
\end{eqnarray}
From \eqref{L1}, since $R=[J,P]$, we have
\begin{eqnarray}\label{P}
 \tilde{P}=P-\theta \Lambda \theta^{-1}
\end{eqnarray}
which can be written in a quasideterminant structure as
\begin{eqnarray}
 \tilde{P}=P+\left|\begin{array}{cc} \theta & I_2\\ \theta\Lambda & \boxed{0_2}\end{array}\right|.
\end{eqnarray}
We rewrite \eqref{P} as
\begin{eqnarray}
 P_{[2]}=P_{[1]}-\theta_{[1]} \Lambda_1 \theta_{[1]}^{-1}
\end{eqnarray}
where $P_{[1]}=P$, $P_{[2]}=\tilde{P}$, $\theta_{[1]}=\theta_1=\theta$, $\Lambda_1=\Lambda$ and $\lambda=\lambda_1$.
Then after $n$ repeated Darboux transformations, we have
\begin{eqnarray}
 P_{[n+1]}=P_{[n]}-\theta_{[n]} \Lambda_n \theta_{[n]}^{-1}
\end{eqnarray}
in which $\theta_{[k]}=\phi_{[k]}\mid_{\phi\rightarrow \theta_k}$. We express $P_{[n+1]}$ in quasideterminant form as
\begin{eqnarray}\label{Pn1}
 P_{[n+1]}=P+\left|\begin{array}{ccccc}
      \theta_1 & \theta_2 \hspace{0.2cm} \ldots \hspace{0.2cm} \theta_n & 0_2\\
      \theta_1\Lambda_1 & \theta_2\Lambda_2 \ldots \theta_n\Lambda_n & 0_2\\
      \vdots & \vdots \hspace{0.3cm} \ldots \hspace{0.3cm} \vdots & \vdots \\
      \theta_1\Lambda_1^{n-2} & \theta_2\Lambda_2^{n-2} \ldots \theta_n\Lambda_n^{n-2} & 0_2\\
      \theta_1\Lambda_1^{n-1} & \theta_2\Lambda_2^{n-1} \ldots \theta_n\Lambda_n^{n-1} & I_2\\
      \theta_1\Lambda_1^n & \theta_2\Lambda_2^n \ldots \theta_n\Lambda_n^n & \boxed{0_2}\\
     \end{array}
 \right|.
\end{eqnarray}
We now express each $\theta_i$, $\Lambda_i$ as a $2 \times 2$ matrix
\begin{eqnarray}
 \theta_i=\left(\begin{array}{cc}
        \phi_{2i-1} & \phi_{2i}\\
        \psi_{2i-1} & \psi_{2i}
  \end{array}\right),~
 \Lambda_i=\left(\begin{array}{cc}
         1 & 0\\
        0 & -1
  \end{array}\right)\lambda_i
\end{eqnarray}
so that
\begin{eqnarray}
 \theta_i\Lambda_i^k=\left(\begin{array}{cc}
        \phi_{2i-1} & (-1)^k\phi_{2i}\\
        \psi_{2i-1} & (-1)^k\psi_{2i}
  \end{array}\right)\lambda_i^k
\end{eqnarray}
for positive integers $i,k=1,\ldots, n$. Here the relation $\phi_{2i-1}\psi_{2i}+\phi_{2i}\psi_{2i-1}=0$ holds.

Let
\begin{eqnarray}
 \Theta^{(n)}=\left(\theta_1\Lambda_1^n,\ldots,\theta_n\Lambda_n^n\right)
       =\left(\begin{array}{c}\phi^{(n)}\\\psi^{(n)} \end{array}\right),
\end{eqnarray}
where
\begin{eqnarray*}
\phi^{(n)}&=&\left(\lambda_1^n\phi_1, \left(-\lambda_1\right)^n\phi_2 , \ldots, \lambda_n^n\phi_{2n-1}, \left(-\lambda_n\right)^n\phi_{2n}\right),\\
\psi^{(n)}&=&\left(\lambda_1^n\psi_1, \left(-\lambda_1\right)^n\psi_2 , \ldots, \lambda_n^n\psi_{2n-1}, \left(-\lambda_n\right)^n\psi_{2n}\right)
\end{eqnarray*}
denote $1 \times 2n$ row vectors. Thus, \eqref{Pn1} can be rewritten as
\begin{eqnarray}
  P_{[n+1]}=P+\left|\begin{array}{cc} \widehat{\Theta} & E\\ \theta^{(n)} & \boxed{0_2}\end{array}\right|,
\end{eqnarray}
where $\widehat{\Theta}=\left(\theta_i\Lambda_i^{j-1}\right)_{i,j=1,\ldots,n}$ and $E=\left(e_{2n-1}, e_{2n}\right)$ denote
$2n \times 2n$ and $2n \times 2$ matrices respectively, where $e_i$ represents a column vector with $1$ in the $i^{th}$ row and zeros elsewhere.
Hence, we obtain
\begin{eqnarray}
 P_{[n+1]}=P+\left(\begin{array}{cc}
 \left|\begin{array}{cc} \widehat{\Theta} & e_{2n-1}\\  \phi^{(n)} & \boxed{0}\end{array}\right| &
 \left|\begin{array}{cc} \widehat{\Theta} & e_{2n}\\  \phi^{(n)} & \boxed{0}\end{array}\right|\\\\
 \left|\begin{array}{cc} \widehat{\Theta} & e_{2n-1}\\  \psi^{(n)} & \boxed{0}\end{array}\right| &
 \left|\begin{array}{cc} \widehat{\Theta} & e_{2n}\\  \psi^{(n)} & \boxed{0}\end{array}\right| \end{array}\right).
\end{eqnarray}
By comparing with \eqref{qr}, we immediately see that $q_{[n+1]}$ and $r_{[n+1]}$ can be expressed as quasideterminants, namely,
\begin{eqnarray}
 q_{[n+1]}=q+2i\left|\begin{array}{cc} \widehat{\Theta} & e_{2n}\\  \phi^{(n)} & \boxed{0}\end{array}\right|,\quad
 r_{[n+1]}=r-2i\left|\begin{array}{cc} \widehat{\Theta} & e_{2n-1}\\  \psi^{(n)} & \boxed{0}\end{array}\right|.
\end{eqnarray}
\[\]
We now consider  the linear eigenvalue problems $L(\Phi_i)=M(\Phi_i)=0$, where the operators $L$, $M$ are given in \eqref{LaxL}-\eqref{LaxM} and $\Phi_i$ denotes $n$ distinct eigenfunctions as
\begin{eqnarray}
  \Phi_i=\left(\begin{array}{c}\phi_i\\\psi_i \end{array}\right) (i=1,\ldots,n).
\end{eqnarray}
Thus, the pair $q_{[n+1]}$ and $r_{[n+1]}$  are written with respect to $n$, where $n$ is an odd $(n=2k-1)$ or even number $(n=2k)$, and $k\in \mathbb{N}$ is a positive integer.
\\[0.2cm]
\textit{In the case of} $n$ \textit{odd} $(n=2k-1)$
\begin{eqnarray}
q_{[n+1]}=q+2i\left|\begin{array}{ccccc}
      \psi_1 & \psi_2 & \ldots & \psi_n & 0\\
      \phi_1\lambda_1 & \phi_2\lambda_2& \ldots & \phi_n\lambda_n & 0\\
      \psi_1\lambda_1^2 & \psi_2\lambda_2^2& \ldots & \psi_n\lambda_n^2 & 0\\
      \vdots & \vdots &  & \vdots & \vdots \\
      \phi_1\lambda_1^{n-2} & \phi_2\lambda_2^{n-2}& \ldots &\phi_n\lambda_n^{n-2} & 0\\
      \psi_1\lambda_1^{n-1} & \psi_2\lambda_2^{n-1}& \ldots &\psi_n\lambda_n^{n-1} & 1\\
      \phi_1\lambda_1^n & \phi_2\lambda_2^n &\ldots &\phi_n\lambda_n^n & \boxed{0}\\
   \end{array}
 \right|,
\end{eqnarray}
\begin{eqnarray}
r_{[n+1]}=r-2i\left|\begin{array}{ccccc}
      \phi_1 & \phi_2 & \ldots & \phi_n & 0\\
      \psi_1\lambda_1 & \psi_2\lambda_2& \ldots & \psi_n\lambda_n & 0\\
      \phi_1\lambda_1^2 & \phi_2\lambda_2^2& \ldots & \phi_n\lambda_n^2 & 0\\
      \vdots & \vdots &  & \vdots & \vdots \\
      \psi_1\lambda_1^{n-2} & \psi_2\lambda_2^{n-2}& \ldots &\psi_n\lambda_n^{n-2} & 0\\
      \phi_1\lambda_1^{n-1} & \phi_2\lambda_2^{n-1}& \ldots &\phi_n\lambda_n^{n-1} & 1\\
      \psi_1\lambda_1^n & \psi_2\lambda_2^n &\ldots &\psi_n\lambda_n^n & \boxed{0}\\
   \end{array}
 \right|.
\end{eqnarray}
\\[0.2cm]
For $n=1$, we obtain a pair of new solutions for the Gerdjikov-Ivanov system \eqref{GI1}-\eqref{GI2}
\begin{eqnarray}\label{q2}
 \begin{array}{ccc}
  q_{[2]}=q+2i\left|\begin{array}{cc} \psi_1 & 1\\ \phi_1\lambda_1 & \boxed{0}\end{array}\right|\end{array}\nonumber\\
         =q-2i\lambda_1\frac{\phi_1}{\psi_1},\label{qr1}\hspace{1.45cm}
 \end{eqnarray}
\begin{eqnarray}
  \begin{array}{ccc}
  r_{[2]}=r-2i\left|\begin{array}{cc} \phi_1 & 1\\ \psi_1\lambda_1 & \boxed{0} \end{array}\right|\end{array}\nonumber\\
         =r+2i\lambda_1\frac{\psi_1}{\phi_1}\label{qr2},\hspace{1.45cm}
\end{eqnarray}
where $\Phi_1=(\phi_1, \psi_1)^T$ is a solution of the eigenvalue problems $L(\Phi_1)=M(\Phi_1)=0$.
$\\[0.2cm]$
\textit{In the case of} $n$ \textit{even} $(n=2k)$
\begin{eqnarray}
q_{[n+1]}=q+2i\left|\begin{array}{ccccc}
      \phi_1 & \phi_2 & \ldots & \phi_n & 0\\
      \psi_1\lambda_1 & \psi_2\lambda_2& \ldots & \psi_n\lambda_n & 0\\
      \phi_1\lambda_1^2 & \phi_2\lambda_2^2& \ldots & \phi_n\lambda_n^2 & 0\\
      \vdots & \vdots &  & \vdots & \vdots \\
      \phi_1\lambda_1^{n-2} & \phi_2\lambda_2^{n-2}& \ldots &\phi_n\lambda_n^{n-2} & 0\\
      \psi_1\lambda_1^{n-1} & \psi_2\lambda_2^{n-1}& \ldots &\psi_n\lambda_n^{n-1} & 1\\
      \phi_1\lambda_1^n & \phi_2\lambda_2^n &\ldots &\phi_n\lambda_n^n & \boxed{0}\\
   \end{array}
 \right|,
\end{eqnarray}
\begin{eqnarray}
r_{[n+1]}=r-2i\left|\begin{array}{ccccc}
      \psi_1 & \psi_2 & \ldots & \psi_n & 0\\
      \phi_1\lambda_1 & \phi_2\lambda_2& \ldots & \phi_n\lambda_n & 0\\
      \psi_1\lambda_1^2 & \psi_2\lambda_2^2& \ldots & \psi_n\lambda_n^2 & 0\\
      \vdots & \vdots &  & \vdots & \vdots \\
      \psi_1\lambda_1^{n-2} & \psi_2\lambda_2^{n-2}& \ldots &\psi_n\lambda_n^{n-2} & 0\\
      \phi_1\lambda_1^{n-1} & \phi_2\lambda_2^{n-1}& \ldots &\phi_n\lambda_n^{n-1} & 1\\
      \psi_1\lambda_1^n & \psi_2\lambda_2^n &\ldots &\psi_n\lambda_n^n & \boxed{0}\\
   \end{array}
 \right|.
\end{eqnarray}
For $n=2$, we have
\begin{eqnarray}
 \begin{array}{ccc}
  q_{[3]}=q+2i\left|\begin{array}{ccc} \phi_1 & \phi_2 & 0\\ \psi_1\lambda_1 & \psi_2 \lambda_2 & 1 \\ \phi_1\lambda_1^2 & \phi_2\lambda_2^2 & \boxed{0}
  \end{array}\right|\end{array},
\end{eqnarray}
\begin{eqnarray}
 \begin{array}{ccc}
  r_{[3]}=r-2i\left|\begin{array}{ccc} \psi_1 & \psi_2 & 0\\ \phi_1\lambda_1 & \phi_2\lambda_2 & 1 \\ \psi_1\lambda_1^2 & \psi_2\lambda_2^2 & \boxed{0}
  \end{array}\right|\end{array}.
\end{eqnarray}
Thus, we obtain a pair of new solutions for the system \eqref{GI1}-\eqref{GI2}, namely
\begin{eqnarray}
 q_{[3]}&=&q-2i\left(\lambda_1^2-\lambda_2^2\right)\frac{\phi_1\phi_2}{\lambda_1\psi_1\phi_2-\lambda_2\phi_1\psi_2},\label{q3}\\
 r_{[3]}&=&r+2i\left(\lambda_1^2-\lambda_2^2\right)\frac{\psi_1\psi_2}{\lambda_1\phi_1\psi_2-\lambda_2\psi_1\phi_2},
\end{eqnarray}
where $\Phi_i=(\phi_i, \psi_i)^T$ is a solution of the eigenvalue problems $L(\Phi_i)=M(\Phi_i)=0~~ (i=1, 2)$.

\subsection*{\textit{Reduction}}
The eigenfunction $\Phi_k=\left(\begin{array}{c}\phi_k\\\psi_k \end{array}\right)$ associated with the eigenvalue $\lambda_k$ has the following relations when we choose the reduction $r_{[n+1]}=q_{[n+1]}^*:$
\begin{eqnarray}
 \psi_k&=&\phi_k^*\hspace{0.2cm}\text{for real}\hspace{0.2cm}\lambda_k, \label{con1}\\
 \psi_k&=&\phi_l^*\hspace{0.2cm}\text{when}\hspace{0.2cm}\lambda_k=\lambda_l^*\ (k\neq l),\label{con2}
\end{eqnarray}
where $k\in \mathbb{N}$. There are many ways to guarantee the reduction $r_{[n+1]} = q_{[n+1]}^*$ for the $n$-fold Darboux transformations when $n>2$.
In the present paper we will restrict ourselves to the reductions $r_{[2]}=q_{[2]}^*$ and $r_{[3]}=q_{[3]}^*$. For the one-fold Darboux transformation, the reduction $r_{[2]}=q_{[2]}^*$ implies
\begin{eqnarray}\label{r1}
 \psi_1=\phi_1^*\hspace{0.2cm} \text{for}\hspace{0.2cm}\lambda_1\in \mathbb{R}.
\end{eqnarray}
Furthermore, for the two-fold Darboux transformation, in order that $r_{[3]}=q_{[3]}^*$, the eigenfunctions $\Phi_1=(\phi_1, \psi_1)^T$ and $\Phi_2=(\phi_2, \psi_2)^T$ with the eigenvalues $\lambda_1, \lambda_2$, either of the following conditions hold:
\begin{eqnarray}
  &&\psi_1=\phi_1^*,~\psi_2=\phi_2^*\hspace{0.2cm}\text{for}\hspace{0.2cm}\lambda_1,\lambda_2 \in \mathbb{R}~\text{or}\label{R1}\\
  &&\psi_1=\phi_2^*,~\psi_2=\phi_1^*\hspace{0.2cm}\text{for}\hspace{0.2cm}\lambda_2^*=\lambda_1 \label{R2}.
\end{eqnarray}

\section{Particular solutions}\label{ParS}

Let us consider the spectral problem $L(\Phi)=M(\Phi)=0$ with eigenvalue $\lambda$, where $\Phi=(\phi, \psi)^T$ and $L, M$ are given by  \eqref{LaxL}-\eqref{LaxM} so that \begin{eqnarray}
 \Phi_x+J\Phi\lambda^2-R\Phi\lambda+\frac{1}{2}qrJ\Phi&=&0\label{Laxeqnx},\\
 \Phi_t+2J\Phi\lambda^4-2R\Phi\lambda^3+qrJ\Phi\lambda^2+U\Phi\lambda+W\Phi&=&0\label{Laxeqnt}.
\end{eqnarray}

\subsection{Solutions for the vacuum}

For $q=r=0$, the above equations transform into the first-order linear system
\begin{eqnarray}
 \Phi_x+J\Phi\lambda^2&=&0\\
 \Phi_t+2J\Phi\lambda^4&=&0
\end{eqnarray}
which has solution
\begin{eqnarray}\label{phi}
 \phi_k=e^{-i\lambda_k^2\left(x+2\lambda_k^2t\right)},\hspace{0.2cm}\psi_k=e^{i\lambda_k^2\left(x+2\lambda_k^2t\right)},
\end{eqnarray}
where $k\in\mathbb{N}$.

\subsubsection*{Case 1 ($n=1$)}

For one single Darboux transformation, due to the required reduction $r=q^*$, we must take $\lambda_1$ to be real and $\psi_1=\phi_1^*$.
By substituting $\phi_1=e^{-i\lambda_1^2\left(x+2\lambda_1^2t\right)}$ and $\psi_1=e^{i\lambda_1^2\left(x+2\lambda_1^2t\right)}$ into \eqref{qr1}, we obtain a new solution $q_{[2]}$ for the GI equation \eqref{GIeqn} as
\begin{eqnarray}
 q_{[2]}=-2i\lambda_1e^{-2i\lambda_1^2\left(x+2\lambda_1^2t\right)},
\end{eqnarray}
where $r_{[2]}=q_{[2]}^*$. This, of course, is not a soliton but a periodic solution. It is obvious that $|q_{[2]}|^2=constant$ so that it satisfies a linear equation $iq_t+q_{xx}=0$ obtained from \eqref{GIeqn}. However, this is not an interesting solution obtained by the use of the Darboux transformation.

\subsubsection*{Case 2 ($n=2$)}

In order that $r_{[3]}=q_{[3]}^*$, $\lambda_1$ and $\lambda_2$ are either real  or
complex conjugate eigenvalues to each other.

\subsubsection*{Case 2a}Under the condition \eqref{R1}, \eqref{q3} yields a periodic solution
\begin{eqnarray}
 q_{[3]}=-2i\frac{\lambda_1^2-\lambda_2^2}{\lambda_1 e^{2i\lambda_1^2\left(x+2\lambda_1^2t\right)}-\lambda_2 e^{2i\lambda_2^2\left(x+2\lambda_2^2t\right)}}
\end{eqnarray}
which can be rewritten as
\begin{eqnarray}\label{parsol}
 \left|q_{[3]}\right|^2=4 \frac{\left(\lambda_1^2-\lambda_2^2\right)^2}
{\lambda_1^2+\lambda_2^2-2\lambda_1\lambda_2 \cos \gamma},
\end{eqnarray}
where $\gamma=2\left(\lambda_1^2-\lambda_2^2\right)\left[x+2(\lambda_1^2+\lambda_2^2)t\right]$.  Here, it can be easily seen that the denominator of the function $\left|q_{[3]}\right|^2$ is positive since $0< \left(\lambda_{1}-\lambda_{2}\right)^{2}\leq \lambda_{1}^{2}+\lambda_{2}^{2}-\lambda_{1}\lambda_{2}\cos\gamma \leq  \left(\lambda_{1}+\lambda_{2}\right)^{2}$. The solution \eqref{parsol} is plotted in the figure 1.
\begin{figure}
\begin{center}
 \includegraphics[width=.4\textwidth]{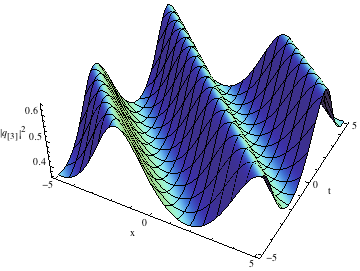}
\center{\footnotesize FIGURE 1. Periodic solution $|q_{[3]}|^{2}$  of the GI equation \eqref{GIeqn} with $\lambda_{1}=0.1, \lambda_{2}=0.7$.}
\end{center}
\end{figure}
\subsubsection*{Case 2b} For the choice \eqref{R2},
\eqref{q3}  leads to
\begin{eqnarray}
 q_{[3]}=-2i\frac{\lambda_1^2-\lambda_2^2}{\lambda_1e^{2i\lambda_1^2\left(x+2\lambda_1^2t\right)}-\lambda_2e^{2i\lambda_2^2\left(x+2\lambda_2^2t\right)}}.
\end{eqnarray}
By taking $\lambda_1=\xi+i\eta$ and $\lambda_2=\xi-i\eta$, we obtain one soliton solution of the GI equation \eqref{GIeqn} as
\begin{eqnarray}
 \left|q_{[3]}\right|^2=32~\frac{\xi^2\eta^2}{\eta^2-\xi^2+\left(\xi^2+\eta^2\right)\cosh\left(8\xi\eta\left[x+4\left(\xi^2-\eta^2\right)\right]\right)},
\end{eqnarray}
where $\xi, \eta \in \mathbb{R}$. The figure 2 demonstrates one soliton solution.
\begin{figure}
\begin{center}
 \includegraphics[width=.4\textwidth]{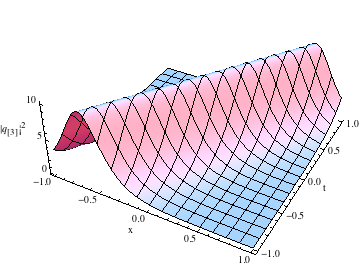}
\center{\footnotesize FIGURE 2. One soliton solution $|q_{[3]}|^{2}$ of the GI equation \eqref{GIeqn} when $\xi=0.8, \eta=0.9$.}
\end{center}
\end{figure}
\subsection{Solutions for non-zero seeds}

For $q, r\neq 0$ and $r=q^*$, it is easily seen that
\begin{eqnarray}\label{qseed}
 q=ke^{i\left[ax+\left(ak^2+\frac{1}{2}k^4-a^2\right)t\right]}
\end{eqnarray}
is a periodic solution of the GI equation \eqref{GIeqn}, where $a$ and $k$ are real numbers.
We use this as the seed solution for application of Darboux transformations.

Substituting \eqref{qseed} into the linear system \eqref{Laxeqnx}-\eqref{Laxeqnt} and then solving for the eigenfunction $\Phi=(\phi, \psi)^T$, we obtain
\begin{eqnarray}
 \phi(x,t,\lambda)&=&c_1e^{\frac{i}{2}\left(\left[a+D\right]x+\left[b-\left(a-2\lambda^2\right)D\right]t\right)}+
c_2e^{\frac{i}{2}\left(\left[a-D\right]x+\left[b+\left(a-2\lambda^2\right)D\right]t\right)},\\
\psi(x,t,\lambda)&=&\widetilde{c_1}e^{-\frac{i}{2}\left(\left[a+D\right]x+\left[b-\left(a-2\lambda^2\right)D\right]t\right)}+
\widetilde{c_2}e^{-\frac{i}{2}\left(\left[a-D\right]x+\left[b+\left(a-2\lambda^2\right)D\right]t\right)},
\end{eqnarray}
where $b=ak^2-a^2+\frac{k^4}{2},~$ $D=\sqrt{a^2+4a\lambda^2+4\lambda^4+k^4+2ak^2},~$ $\widetilde{c_1}=i\left(\frac{k^2+a+2\lambda^2-D}{2k\lambda}\right)c_2,~$
$\widetilde{c_2}=i\left(\frac{k^2+a+2\lambda^2+D}{2k\lambda}\right)c_1$ and $c_1$, $c_2$ are integration constants, obtained from \eqref{Laxeqnx}-\eqref{Laxeqnt}.
\vspace{1 mm}

\subsubsection*{Case 3 ($n=1$)}

For the one-fold Darboux transformation, it can easily be shown that $D^2(\lambda_1)>0$ and
$D^2(\lambda_1)<0$ produce the periodic and soliton solutions respectively of the GI equation. 
For example, for $D^2(\lambda_1)>0$ with $k^{2}=-2a$, \eqref{q2} yields a periodic solution
\begin{eqnarray}\label{persolq2}
 \left|q_{[2]}\right|^2=2 \frac{\left(a+2\lambda_1^2\right)^2}
{2\lambda_1^2-a-2k\lambda_1 \sin \gamma},
\end{eqnarray}
where $\gamma=\left(a+2\lambda_1^2\right)x+(4\lambda_1^{4}-a^2)t$. In this solution, it should be observed that  the denominator  must be positive since 
$2\lambda_1^2-a-2k\lambda_1 \sin \gamma=\frac{1}{2}\left(4\lambda_1^2+k^{2}-4k\lambda_1 \sin \gamma\right)\geq \frac{1}{2}\left(2\lambda_{1}-k\right)^{2}>0$ for $k\neq2\lambda_{1}$.
The solution \eqref{persolq2} is plotted in the figure 3.
\begin{figure}
\begin{center}
 \includegraphics[width=.4\textwidth]{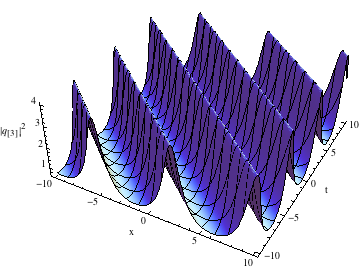}
\center{\footnotesize FIGURE 3. Periodic solution $|q_{[2]}|^{2}$  of the GI equation \eqref{GIeqn} with the choice of parameters
$\lambda_{1}=0.3, k=\sqrt{2}$.}
\end{center}
\end{figure}

\subsubsection*{Case 4 ($n=2$)}

In this case, we have two eigenvalues $\lambda_1$ and $\lambda_2$.  For solutions such that $r_{[3]}=q_{[3]}^{*}$, these eigenvalues are either real or complex conjugate to each other. The functions $\phi_{i}, \psi_{i} (i=1,2)$ with  the eigenvalues $\lambda_{1}$, $\lambda_{2}$ either hold $(R1)\; \psi_{1}=\phi_{1}^{*}, \psi_{2}=\phi_{2}^{*}$ for $\lambda_{1},\lambda_{2}\in\mathbb{R}$ or $(R2)\; \psi_{1}=\phi_{2}^{*}, \psi_{2}=\phi_{1}^{*}$ for $\lambda_{1}=\lambda_{2}^{*}$. An example for ($R2$) is given below.

The solution \eqref{q3} under the choice ($R2$) can be rewritten as
\begin{eqnarray}
 q_{[3]}=q+2i\Lambda\frac{\phi_1\phi_2}{\lambda_2\left|\phi_{1}\right|^{2}-\lambda_1\left|\phi_2\right|^{2}},
\end{eqnarray}
where $\Lambda=\lambda_1^2-\lambda_2^2$ such that $\Lambda \in i\mathbb{R}$ . 
For simplicity, let us choose $k^{2}=-2a$, then we find 
\begin{eqnarray}\label{q3.R2}
 q_{[3]}=\frac{k\lambda_{2}\Lambda_{2} e^{i\left([a+\Lambda]x-\left[a^{2}-2\kappa\Lambda\right]t\right)}
              -k\lambda_{1}\Lambda_{1} e^{i\left(\left[a-\Lambda\right]x-\left[a^{2}+2\kappa\Lambda\right]t\right)}
             -4i\lambda_{1}\lambda_{2}\Lambda e^{-i\left(\kappa x+2\left[\lambda_{1}^{4}+\lambda_{2}^{4}\right]t\right)}}
            { \lambda_{2}\Lambda_{1} e^{i\Lambda\left(x+2\kappa t\right)}-
              \lambda_{1}\Lambda_{2} e^{-i\Lambda\left(x+2\kappa t\right)}              
               -ik\Lambda e^{i\left(\left[a+\kappa\right]x-\left[a^{2}-2\lambda_{1}^{4}-2\lambda_{2}^{4}\right]t\right)}},
 \end{eqnarray}
where  $\Lambda_{1}=a+2\lambda_1^2$, $\Lambda_{2}=a+2\lambda_2^2$ and $\kappa=\lambda_{1}^{2}+\lambda_{2}^{2}$ such that $\kappa\in \mathbb{R}$.

\vspace{2mm}

Let  $\lambda_{1}=\xi+i\eta$ and $\lambda_{2}=\xi-i\eta$ , where $\xi,\eta \neq 0$.  Then
\begin{eqnarray}
 \left|q_{[3]}\right|^{2}=k^{2}+16 \xi \eta ~\frac{m_{0}+m_{1}\cos\gamma_{1}\sinh\gamma_{2}+m_{2}\sin\gamma_{1}\cosh\gamma_{2}}
 {n_{0}+n_{1}\cosh(2\gamma_{2})+n_{2}\cos\gamma_{1}\sinh\gamma_{2}+n_{3}\sin\gamma_{1}\cosh\gamma_{2}},
\end{eqnarray}
where
\begin{eqnarray*}
  \gamma_{1}&=&\left(a+2\left[\xi^{2}-\eta^{2}\right]\right)x-\left(a^{2}-4\left[\xi^{2}-\eta^{2}\right]^{2}+16\xi^{2}\eta^{2}\right)t,\\
  \gamma_{2}&=&4\xi\eta\left(x+4\left[\xi^{2}-\eta^{2}\right]t\right),\\
  m_{0}&=&2\xi\eta\left(3a^{2}+4a\left[\xi^{2}-\eta^{2}\right]-4\left[\xi^{2}+\eta^{2}\right]^{2}\right),\\
  m_{1}&=&k\xi\left(a^{2}+4\left[\xi^{2}+\eta^{2}\right]\left[a+\xi^{2}-3\eta^{2}\right]\right),\\    
  m_{2}&=&k\xi\left(a^{2}-4\left[\xi^{2}+\eta^{2}\right]\left[a+3\xi^{2}-\eta^{2}\right]\right),\\   
  n_{0}&=&\left(\xi^{2}-\eta^{2}\right)\left(a+2\xi^{2}+2\eta^{2}\right)^{2}+8a\eta^{2}\left(3\xi^{2}+\eta^{2}\right),\\
  n_{1}&=&-\left(\xi^{2}+\eta^{2}\right)\left(\left[a+2\xi^{2}+2\eta^{2}\right]^{2}-8a\eta^{2}\right),\\
  n_{2}&=&8k\xi^{2}\eta\left(a+2\xi^{2}+2\eta^{2}\right),\\
  n_{3}&=&8k\xi^{2}\eta^{2}\left(a-2\xi^{2}-2\eta^{2}\right).
\end{eqnarray*}
By choosing appropriate parameters, the breather solution of the Gerdijikov-Ivanov equation \eqref{GIeqn} is plotted in the figure 4.
\begin{figure}
\begin{center}
 \includegraphics[width=.4\textwidth]{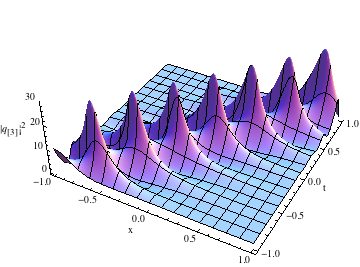}
\center{\footnotesize FIGURE 4. Breather solution $|q_{[3]}|^{2}$  of the GI equation \eqref{GIeqn} with the parameters chosen as  $\xi=1, \eta=1.1, k=\sqrt{2}$.}
\end{center}
\end{figure}
Similarly, for the choice ($R1$), \eqref{q3} gives us a periodic solution.

\section{Conclusion}\label{Con}

In this paper, we have presented a standard Darboux transformation for the GI equation \eqref{GIeqn}. We have constructed solutions in quasideterminant forms for the GI equation. These quasideterminants  are expressed in terms of  solutions of  the linear partial differential equations given by  \eqref{Laxeqnx}-\eqref{Laxeqnt}. These solutions arise naturally from the Darboux transformation we present here.  Moreover, parametric, soliton and breather solutions for zero and non-zero seeds have been given as particular examples for the GI equation. Examples of these particular solutions are plotted in the figures 1--4 with the chosen parameters.
It should be emphasised that we may derive several types of particular solutions for the GI equation by using the Darboux transformation we present here. Finally, it  should be pointed out that the Darboux transformation technique is a universal instrument that allows us to construct exact solutions for other integrable systems.

\subsection*{Acknowledgments}The author would like to thank Dr. Jonathan Nimmo for his valuable comments and suggestions.


\begin{thebibliography}{99}
\bibitem{Agr}G.P. Agrawal, Nonlinear Fibers Optics (Academic, New York, 2001).
\bibitem{Chen}H.H. Chen, Y.C. Lee and C.S. Liu, Integrability of Nonlinear Hamiltonian Systems by Inverse Scattering Method, Phys. Scr. \textbf{20} (1979) 490--492.
\bibitem{Clarkson}P.A. Clarkson and J.A. Tuszynski, Exact solutions of the multidimensional derivative nonlinear Schr\"{o}dinger equation for many-body systems near criticality, J. Phys. A: Math. Gen. \textbf{23} (1990) 4269--4288.
\bibitem{Clarkson87}P.A. Clarkson and C.M. Cosgrove, PainlevŽ analysis of the nonlinear Schr\"{o}dinger family of equations, J. Phys. A \textbf{20}(8) (1987) 2003Ð2024. 
\bibitem{Darboux} G. Darboux, Comptes Rendus de l'Acadmie des Sciences \textbf{94} (1882) 1456--1459.
\bibitem{Fan} E. Fan, Darboux transformation and soliton-like solutions for the Gerdjikov--Ivanov equation, J. Phys.A: Math. Gen. \textbf{33} (2000) 6925--6933.
\bibitem{Gelfand91}I. Gelfand and V. Retakh, Determinants of the matrices over noncomutative rings, Funct. Anal. App. \textbf{25} (1991) 91--102.
\bibitem{Gelfand05}I. Gelfand, S. Gelfand, V. Retakh, and R. L. Wilson. Quasideterminants. Adv.Math. \textbf{193}(1) (2005) 56--141.
\bibitem{CJ}C.R. Gilson and J.J.C. Nimmo, On a direct approach to quasideterminant solutions of a noncommutative KP equation,
              J. Phys. A: Math. Theor. \textbf{40}(14) (2007) 3839--3850.
\bibitem{GNH} C.R. Gilson, M. Hamanaka and J.C. Nimmo,
              B\"{a}cklund transformations and the Atiyah-Ward ansatz for non-commutative anti-self-dual Yang-Mills equations,
              Proc. R. Soc. A \textbf{465} (2009) 2613--2632.
\bibitem{GI}V.S. Gerdjikov and M.I. Ivanov, A quadratic pencil of general type and nonlinear evolution equations. II. Hierarchies of Hamiltonian structures, Bulgar. J. Phys. \textbf{10} (1983) 130--143.
\bibitem{Guo}L. Guo, Y. Zhang, S. Xu, W. Zhiwei and J. He, The higher order rogue wave solutions of the GerdjikovÐIvanov equation, Phys. Scr. \textbf{89} (2014) 035501.
\bibitem{Hassan}M. Hassan, Darboux transformation of the generalized coupled dispersionless integrable system, J.Phys.A: Math. Theor. \textbf{42} (2009) 065203.
\bibitem{Ich} Y.H. Ichikawa, K. Konno, M. Wadati and H. Sanuki, Spiky soliton in circular polarized Alfv$\acute{e}$n wave, J. Phys.Soc. Japan \textbf{48} (1980) 279--286.
\bibitem{Johnson} R.S. Johnson, On the modulation of water waves in the neighbourhood of $\it{kh} \approx$ 1.363, Proc. Roy.Soc.London Ser.A \textbf{357} (1977) 131--141.
\bibitem{KN} D.J. Kaup and A.C. Newell, An exact solution for derivative nonlinear Schr\"{o}dinger equation, J.Math.Phys. \textbf{19}(4) (1978) 798--801.
\bibitem{Kodama} Y. Kodama, Optical solitons in a monomode fiber, J. Statist. Phys. \textbf{39} (1985) 597--614.
\bibitem{Kakei}S. Kakei, N. Sasa and J. Satsuma, Bilinearization of a generalized derivative nonlinear Schršdinger equation, J. Phys. Soc. Japan \textbf{64}(5) (1995) 1519--1523. 
\bibitem{Kundu84}A. Kundu, Landau-Lifshitz and higher-order nonlinear systems gauge generated from nonlinear Schr\"{o}dinger type equations, J. Math. Phys. \textbf{25}(12) (1984) 3433--3438.
\bibitem{Kundu} A. Kundu, Exact solutions to higher-order nonlinear equations through gauge transformation, Physica \textbf{25}D (1987) 399--406.
\bibitem{Matveev} V.B. Matveev, Darboux transformation and explicit solutions of the Kadomtcev-Petviaschvily equation, depending on functional parameters. Lett. Math. Phys. \textbf{3} (1979) 213--216.
 \bibitem{MS} V.B. Matveev and M.A. Salle, Darboux transformations and solitons (Springer Series in Nonlinear Dynamics, Springer-Verlag, Berlin, 1991).
 \bibitem{Nimmo06}J. J. C. Nimmo, On a non-Abelian Hirota-Miwa equation, J. Phys. A: Math. Gen. \textbf{39} (2006) 5053--5065.
 \bibitem{NH} J.J.C. Nimmo and H. Yilmaz, On Darboux Transformations for the derivative nonlinear Schr\"{o}dinger equation, Journal of Nonlinear Mathematical Physics \textbf{21}(2) (2014) 278--293.
 \bibitem{NGO}J. J. C. Nimmo, C. R. Gilson and Y. Ohta, Applications of Darboux transformations to the self-dual Yang-Mills equations, Theor. Math. Phys. \textbf{122} (2000) 239--246.
 \bibitem{Ohta} Y. Ohta and J. Yang, General high-order rogue waves and their dynamics in the nonlinear Schr\"{o}dinger equation,
              Proc. R. Soc. A \textbf{468} (2012) 1716--1740.         
 \bibitem{Wadati}M. Wadati and K. Sogo, Gauge transformations in soliton theory, J. Phys. Soc. Japan \textbf{52}(2) (1983) 394--398.   
 \bibitem{Xu} S. Xu and J. He, The rogue wave and breather solution of the Gerdjikov-Ivanov equation, J. Math. Phys.\textbf{53} (2012) 6063507.
\end{thebibliography}
\end{document}